\documentclass[5p,numbers,sort&compress]{elsarticle}

\usepackage[english]{babel}
\usepackage{graphicx, color, epsfig}
\usepackage{enumerate}
\usepackage{latexsym}
\usepackage{fancyhdr}
\usepackage{multirow}
\usepackage{amssymb,amsthm}

\newcommand{\ds}{\displaystyle}
\newcommand{\vs}{\vspace}

\newcommand{\dd}{\mathrm{d}}
\newcommand{\ddd}{\mathbf{d}}
\newcommand{\E}{\mathbf{E}}
\newcommand{\e}{\mathrm{e}}
\newcommand{\ee}{\mathrm{e}}

\newcommand{\g}{\mathrm{g}}

\newcommand{\h}{\mathrm{h}}

\newcommand{\mm}{\mathbf{m}}
\newcommand{\nn}{\mathbf{n}}

\newcommand{\rr}{\mathbf{r}}

\newcommand{\lambdabarre}{\!\!\!
\raisebox{0cm}{$\begin{array}{c}
-
\\[-0.4cm]
\lambda
  \end{array}$}\!\!\!}

\begin{document}
\begin{frontmatter}
\title{Some theoretical results on semiconductor spherical quantum dots}

\author[LPTHE,UCP]{B. {\sc Billaud}}
\author[UCP]{T.-T. {\sc Truong}\corref{corresponding_author}}
\ead{truong@u-cergy.fr}
\cortext[corresponding_author]{To whom correspondence should be addressed.}

\address[LPTHE]{Laboratoire de Physique Th\'eorique et Hautes Energies (LPTHE),
\\
CNRS UMR 7589, Universit\'e Pierre et Marie Curie (Paris VI),
\\
4, place Jussieu, F-75252 Paris Cedex 05, France.}

\address[UCP]{Laboratoire de Physique Th\'eorique et Mod\'elisation (LPTM),
\\
CNRS UMR 8089, Universit\'e de Cergy-Pontoise,
\\
2, avenue Adolphe Chauvin, F-95302 Cergy-Pontoise Cedex, France.}

\begin{abstract}
We use an improved version of the standard effective mass approximation model to describe quantum effects in nanometric semiconductor Quantum Dots (QDs). This allows analytic computation of relevant quantities to a very large extent. We obtain, as a function of the QD radius, in precise domains of validity, the QD excitonic ground state energy and its Stark and Lamb shifts. Finally, the Purcell effect in QDs is shown to lead to potential QD-LASER emitting in the range of visible light.

\end{abstract}
\end{frontmatter}
\section{Introduction}
As semiconducting Quantum Dots (QDs) display standard atomic physics properties, they may be thought as giant artificial atoms, with adjustable quantized energy spectra through their sizes and shapes. They are of interest in a wide range of research areas \cite{Kirkstaedter_1994, Yoshie_2004, Trauzettel_2007, Ishibashi_2003, So_2006}. In such structures, Quantum Size Effects (QSE) are characterized by a blue-shift in the semiconductor optical spectrum, due to the increase of the charge carrier confinement energy. Results on QSE in low-dimensional semiconductor structures and modern approaches to this problem are discussed in \cite{Yoffe_2002, Delerue}. In this paper, the QD problem is dealt within a modified effective mass approximation (EMA) model, to which a pseudo-potential is added ({\it cf.} section {\bf \ref{sec_2}}). This partially removes the over-estimation of the electron-hole pair confinement energy for small QDs, and allows the analytic determination of the Kayanuma function $\eta(\lambda)$ \cite{Kayanuma_1988}.

The physics of QDs, particularly in regard to the QD interaction with an external field, is very attractive. It gives rise to quantum-confinement Stark effects (QCSE) \cite{Nomura_1990, Nomura_1990b}, which manifest themselves through a red-shift of exciton photoluminescence \cite{Yakimov_2003}. In section {\bf \ref{sec_3}}, we use the EMA model to obtain analytic criterions on the QD radius and the applied electric field amplitude, as a result of the interplay between electron-hole Coulomb interaction and an additional polarization energy \cite{Billaud_2009}.  When the electromagnetic field is quantized, an energy level Lamb shift occurs \cite{Bethe_1947, Welton_1948}. While it is a continual subject of research \cite{Sajeev_1990, Rotkin_2000}, it seems to be unknown for QDs. In section {\bf \ref{sec_4}}, we use the EMA framework to uncover an observable negative Lamb shift for the electron-hole pair ground state, in judiciously chosen QDs. In section {\bf \ref{sec_5}}, the Purcell effect, a test bed for many applications \cite{Nakwaski_2003, Santori_2004, Kiraz_2004}, is studied for QDs. A condition for its occurrence, despite the action of unfavorable Rabi oscillations, is derived. This opens the way for a visible light QD-LASER, driven by a Purcell effect.
\section{Quantum Size Effects} \label{sec_2}
In a standard EMA model, an electron and a hole, of effective masses $m^*_{\e,\h}$, behave as free particles in a spherical infinite potential well $V(\rr_{\e,\h})$. Their Coulomb interaction $-\frac{e^2}{\kappa r_{\e\h}}$ is treated by a variational procedure. To handle the interplay between confinement energy (scaling as $\propto R^{-2}$), and Coulomb potential (scaling as $\propto R^{-1}$), two regimes are singled out by the values of the ratio of the QD radius $R$ to the bulk Mott-Wannier exciton Bohr radius $a^*$ \cite{Kayanuma_1988}. Here, as far as Stark, Lamb or Purcell effects are concerned, the so-called weak field limit, by which the charge carriers cannot overstep the real confining potential by tunneling, is assumed.

In a strong confinement regime, where $R\lesssim 2a^*$, the electron-hole relative motion is affected by the infinite potential well, such that the electron-hole pair stays quasi-uncorrelated. To exhibit the excitonic behavior of the electron-hole pair, both electron and hole, individually confined and being in their ground state, the following trial function $\phi(\rr_\e,\rr_\h)\propto\ee^{-\frac\sigma2\frac{r_{\textrm{\tiny eh}}}{a^*}}$, of parameter $\sigma$, is to be used. The electron-hole pair ground state energy $E^\mathrm{strong}_{\e\h}=E_{\e\h}-1.786\frac{e^2}{\kappa R}-0.248E^*$, given in \cite{Kayanuma_1988}, is retrieved up to $(\frac R{a^*})^2$ order, $\mu$ being the electron-hole pair reduced mass, $E_{\e\h}=\frac{\pi^2}{2\mu R^2}$ its ground state confinement energy, and $E^*$ the excitonic Rydberg energy.

In the weak confinement regime, where $R\gtrsim 4a^*$, the exciton behaves as a quasi-particle of mass $M=m^*_\e+m^*_\h$. Then, the leading contribution to its ground state energy is $-E^*$, while its total translational motion is $\frac{\pi^2}{2MR^2}$. To improve the accuracy of the exciton ground state energy $-E^*+\frac{\pi^2}{2MR^2}$, a phenomenological function $\eta(\lambda)$ of the mass ratio $\lambda=\frac{m_\textrm{\tiny h}^*}{m_\textrm{\tiny e}^*}$ was introduced in \cite{Kayanuma_1988}. The exciton is thought as a rigid sphere of radius $\eta(\lambda)a^*$, and its center-of-mass cannot reach the infinite potential well boundary unless the electron-hole relative motion undergoes a strong deformation \cite{Kayanuma_1988}. Thus, adding a ground state plane wave in the center-of-mass coordinates to the trial function $\phi(\rr_\e,\rr_\h)$, we obtain, up to $(\frac{a^*}R)^{3}$ order, $E^{\textrm{\scriptsize weak}}_{\e\h}=-E^*+\frac{\pi^2}{6\mu R^2}+\frac{\pi^2}{2M(R-\eta(\lambda)a^*)^2}$, where $\eta(\lambda)=0.208\frac{(1+\lambda)^2}{\lambda}$. $\eta(\lambda)$ satisfies the electron-hole exchange symmetry, and table \ref{table_1} shows good agreement with computational results.

\begin{table}
\caption{\label{table_1}Comparison of $\eta(\lambda)$ values from computational results of \cite{Kayanuma_1988} and theoretical results  given by equation $\eta(\lambda)=0.208\frac{(1+\lambda)^2}{\lambda}$.} \vs{-.2cm}
    \begin{center}
{\footnotesize
      \begin{tabular}{cccc}
\hline
$\lambda$ & 1 & 3 & 5
\\
\hline
$\eta_{\textrm{\scriptsize comp}}(\lambda)$ & 0.73 & 1.1 & 1.4
\\
$\eta_{\textrm{\scriptsize theo}}(\lambda)$ & 0.83 & 1.1 & 1.5
\\
relative error & $\approx$14\% & $<$1\% & $\approx$7\%
\\
\hline
      \end{tabular}}
    \end{center}
\vs{-.8cm}
\end{table}

\begin{figure}
\caption{Electron-hole pair ground state energy as a function of the QD radius computed for a confining infinite potential well with (--$\!~$--) or without (---) the presence of the pseudo-potential $W(\rr_{\e,\h})$ and for a confining finite potential step of height $V_0\approx1$eV (--$~\!\cdot\!~$--) \cite{Thoai_1990} and compared to experimental results for $CdS$ microcrystallites \cite{Weller_1986}.} \label{figure_1}
\vs{-.2cm}
  \begin{center}
\begin{picture}(0,0)%
\includegraphics{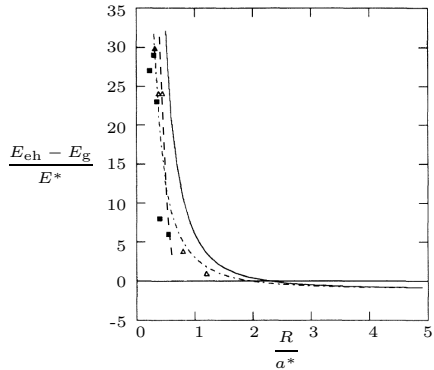}%
\end{picture}%
\setlength{\unitlength}{4144sp}%
\begingroup\makeatletter\ifx\SetFigFontNFSS\undefined%
\gdef\SetFigFontNFSS#1#2#3#4#5{%
  \reset@font\fontsize{#1}{#2pt}%
  \fontfamily{#3}\fontseries{#4}\fontshape{#5}%
  \selectfont}%
\fi\endgroup%
\begin{picture}(2027,2141)(-239,-1259)
\put(-44,-1054){\makebox(0,0)[rb]{\smash{{\SetFigFontNFSS{6}{7.2}{\rmdefault}{\mddefault}{\updefault}{\color[rgb]{0,0,0}-5}%
}}}}
\put( 11,-1109){\makebox(0,0)[lb]{\smash{{\SetFigFontNFSS{6}{7.2}{\rmdefault}{\mddefault}{\updefault}{\color[rgb]{0,0,0}0}%
}}}}
\put(361,-1109){\makebox(0,0)[lb]{\smash{{\SetFigFontNFSS{6}{7.2}{\rmdefault}{\mddefault}{\updefault}{\color[rgb]{0,0,0}1}%
}}}}
\put(685,-1109){\makebox(0,0)[lb]{\smash{{\SetFigFontNFSS{6}{7.2}{\rmdefault}{\mddefault}{\updefault}{\color[rgb]{0,0,0}2}%
}}}}
\put(1037,-1109){\makebox(0,0)[lb]{\smash{{\SetFigFontNFSS{6}{7.2}{\rmdefault}{\mddefault}{\updefault}{\color[rgb]{0,0,0}3}%
}}}}
\put(1713,-1109){\makebox(0,0)[lb]{\smash{{\SetFigFontNFSS{6}{7.2}{\rmdefault}{\mddefault}{\updefault}{\color[rgb]{0,0,0}5}%
}}}}
\put(1388,-1109){\makebox(0,0)[lb]{\smash{{\SetFigFontNFSS{6}{7.2}{\rmdefault}{\mddefault}{\updefault}{\color[rgb]{0,0,0}4}%
}}}}
\put(-44,-596){\makebox(0,0)[rb]{\smash{{\SetFigFontNFSS{6}{7.2}{\rmdefault}{\mddefault}{\updefault}{\color[rgb]{0,0,0}5}%
}}}}
\put(-44,-352){\makebox(0,0)[rb]{\smash{{\SetFigFontNFSS{6}{7.2}{\rmdefault}{\mddefault}{\updefault}{\color[rgb]{0,0,0}10}%
}}}}
\put(-44,-111){\makebox(0,0)[rb]{\smash{{\SetFigFontNFSS{6}{7.2}{\rmdefault}{\mddefault}{\updefault}{\color[rgb]{0,0,0}15}%
}}}}
\put(-44,105){\makebox(0,0)[rb]{\smash{{\SetFigFontNFSS{6}{7.2}{\rmdefault}{\mddefault}{\updefault}{\color[rgb]{0,0,0}20}%
}}}}
\put(-44,348){\makebox(0,0)[rb]{\smash{{\SetFigFontNFSS{6}{7.2}{\rmdefault}{\mddefault}{\updefault}{\color[rgb]{0,0,0}25}%
}}}}
\put(-44,591){\makebox(0,0)[rb]{\smash{{\SetFigFontNFSS{6}{7.2}{\rmdefault}{\mddefault}{\updefault}{\color[rgb]{0,0,0}30}%
}}}}
\put(-44,807){\makebox(0,0)[rb]{\smash{{\SetFigFontNFSS{6}{7.2}{\rmdefault}{\mddefault}{\updefault}{\color[rgb]{0,0,0}35}%
}}}}
\put(902,-1216){\makebox(0,0)[b]{\smash{{\SetFigFontNFSS{6}{7.2}{\rmdefault}{\mddefault}{\updefault}{\color[rgb]{0,0,0}$\ds\frac{R}{a^*}$}%
}}}}
\put(-224,-111){\makebox(0,0)[rb]{\smash{{\SetFigFontNFSS{6}{7.2}{\rmdefault}{\mddefault}{\updefault}{\color[rgb]{0,0,0}$\ds\frac{E_{\e\h}-E_\g}{E^*}$}%
}}}}
\put(-44,-812){\makebox(0,0)[rb]{\smash{{\SetFigFontNFSS{6}{7.2}{\rmdefault}{\mddefault}{\updefault}{\color[rgb]{0,0,0}0}%
}}}}
\end{picture}%
  \end{center}
\vs{-.8cm}
\end{figure}

However, $E^{\textrm{\scriptsize weak}}_{\e\h}$ has a further kinetic energy term $\frac{\pi^2}{6\mu R^2}$ in the relative coordinates \cite{Kayanuma_1988}.  As the virial theorem should be satisfied in these coordinates, this energy is already contained in the Rydberg energy. To remove this contribution, we propose to add the pseudo-potential $W(\rr_{\e\h})=-\frac{32\pi^2}9E^*\frac{r_{\e\h}^2}{R^2}\ee^{-2\frac{r_{\textrm{\tiny eh}}}{a^*}}$.\footnote{$W(\rr_{\e\h})$ should be attractive at distances $\approx a^*$ to promote excitonic state with typical size around $a^*$, repulsive at short distances to penalize small size excitonic states, and exponentially small at large distances.} This pseudo-potential also decreases the exciton energy by $\approx-19.9E^*$ in the strong confinement regime. Figure \ref{figure_1} shows that the excitonic energy computed with $W(\rr_{\e\h})$ shows a better fit to experimental results for $2R\lesssim a^*$. The divergence for very small QD size still persists as a consequence of the infinite potential well assumption \cite{Thoai_1990}. To improve predictive results in this region, energy expansions may be carried out to a few more orders. But, computations become so involved that the relevance of such an approach can be questioned.
\section{Quantum-confinement Stark effects} \label{sec_3}
An electric field $\E_\dd$, is applied along the $z$-axis of a cartesian coordinates system with its origin at the QD center. Even if the EMA model does not fully describe the QD behavior in the absence of electric field, it can be still used to study QCSE in the weak field limit but should include the dipolar interaction $W_{\e\h}(\rr_{\e\h})=\E_\dd\cdot\ddd_{\e\h}$, $\ddd_{\e\h}$ being the exciton dipole moment. The quantity $eE_\dd R$, where $E_\dd=|\E_\dd|$, is treated as a perturbation. Following \cite{Bastard_1983}, we use here the trial function $\phi(\rr_\e,\rr_\h)$, describing the electric field free electron-hole pair, but with electric field interaction factors $\ee^{\mp\frac{\sigma_{\textrm{\tiny e,h}}}2z_\textrm{\tiny e,h}}$ of parameters $\sigma_{\textrm{\tiny e,h}}$, to account for the spherical shape deformation along $\E_\dd$.

The Stark shift is determined, up to $\!~\frac R{a^*}~\!$ order, as $\Delta E^{\textrm{\scriptsize strong}}_{\textrm{\scriptsize Stark}}=-\Gamma Me^2E_\dd^2R^4\!\left\{1+\Gamma_{\e\h}\frac R{a^*}\right\}\!$, where $\Gamma\approx0.018$ and $\Gamma_{\e\h}$ depends on the semiconductor. The first contribution is the sum of the Stark shifts undergone by the electron and hole ground states. The second contribution expresses the remnant of electron-hole pair states as exciton bound states. As the inside semiconducting QD dielectric constant $\varepsilon$ is larger than the outside insulating matrix one, the polarization energy $P(\rr_\e,\rr_\h)$ introduced in \cite{Brus_1984} is also considered, and its relative role {\it vs.} the Coulomb potential explored. Explicit analytical expressions for any Stark effect quantity, pertaining either to the polarization energy or to its combined effect with Coulomb potential, are given in \cite{Billaud_2009}, and successfully confronted with computational data \cite{Nomura_1990b}.

\begin{figure}
\caption{Stark shift for electron-hole pair as a function of the QD radius, when $E_\dd=12.5$kV.cm$^{-1}$, only including the Coulomb interaction ($\Gamma_{\e\h}\approx-0.163$) up to the zeroth (---) or to the first (--$\!~$--) order, only including the polarization energy ($\Gamma_{\e\h}\approx-0.042$) up to the first order (--$\!~\cdot\!~$--), and including both the Coulomb interaction and the polarization energy ($\Gamma_{\e\h}\approx-0.205$) up to to the first order (--$\!~\cdot\cdot\!~$--), in comparison with results (+) from \cite{Nomura_1990}.} \label{figure_2} \vs{-.2cm}
  \begin{center}
~~~~~~~~~~~~~~~~\begin{picture}(0,0)%
\includegraphics{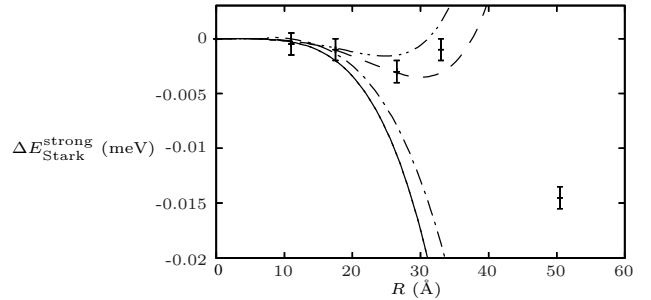}%
\end{picture}%
\setlength{\unitlength}{4144sp}%
\begingroup\makeatletter\ifx\SetFigFontNFSS\undefined%
\gdef\SetFigFontNFSS#1#2#3#4#5{%
  \reset@font\fontsize{#1}{#2pt}%
  \fontfamily{#3}\fontseries{#4}\fontshape{#5}%
  \selectfont}%
\fi\endgroup%
\begin{picture}(2859,1780)(-1726,-902)
\put(-168,-856){\makebox(0,0)[b]{\smash{{\SetFigFontNFSS{6}{7.2}{\rmdefault}{\mddefault}{\updefault}{\color[rgb]{0,0,0}$R$~(\AA)}%
}}}}
\put(-1709,-16){\makebox(0,0)[rb]{\smash{{\SetFigFontNFSS{6}{7.2}{\rmdefault}{\mddefault}{\updefault}{\color[rgb]{0,0,0}$\Delta E^{\textrm{\tiny strong}}_{\textrm{\tiny Stark}}$~(meV)}%
}}}}
\put(-974,-749){\makebox(0,0)[b]{\smash{{\SetFigFontNFSS{6}{7.2}{\rmdefault}{\mddefault}{\updefault} 10}}}}
\put(-571,-749){\makebox(0,0)[b]{\smash{{\SetFigFontNFSS{6}{7.2}{\rmdefault}{\mddefault}{\updefault} 20}}}}
\put(-168,-749){\makebox(0,0)[b]{\smash{{\SetFigFontNFSS{6}{7.2}{\rmdefault}{\mddefault}{\updefault} 30}}}}
\put(235,-749){\makebox(0,0)[b]{\smash{{\SetFigFontNFSS{6}{7.2}{\rmdefault}{\mddefault}{\updefault} 40}}}}
\put(639,-749){\makebox(0,0)[b]{\smash{{\SetFigFontNFSS{6}{7.2}{\rmdefault}{\mddefault}{\updefault} 50}}}}
\put(1042,-749){\makebox(0,0)[b]{\smash{{\SetFigFontNFSS{6}{7.2}{\rmdefault}{\mddefault}{\updefault} 60}}}}
\put(-1404,-688){\makebox(0,0)[rb]{\smash{{\SetFigFontNFSS{6}{7.2}{\rmdefault}{\mddefault}{\updefault}-0.02}}}}
\put(-1404,-350){\makebox(0,0)[rb]{\smash{{\SetFigFontNFSS{6}{7.2}{\rmdefault}{\mddefault}{\updefault}-0.015}}}}
\put(-1404,-10){\makebox(0,0)[rb]{\smash{{\SetFigFontNFSS{6}{7.2}{\rmdefault}{\mddefault}{\updefault}-0.01}}}}
\put(-1404,298){\makebox(0,0)[rb]{\smash{{\SetFigFontNFSS{6}{7.2}{\rmdefault}{\mddefault}{\updefault}-0.005}}}}
\put(-1404,638){\makebox(0,0)[rb]{\smash{{\SetFigFontNFSS{6}{7.2}{\rmdefault}{\mddefault}{\updefault}0}}}}
\end{picture}%
  \end{center}
\vs{-.8cm}
\end{figure}

For $CdS_{0.12}Se_{0.88}$ microcrystals, the strong confinement regime and the weak field limit condition are fulfilled for $E_\dd\approx12.5$kV.cm$^{-1}$ and $R\lesssim30$\AA, if Coulomb interaction and polarization energy are taken to account. When the polarization energy is considered alone, the strong confinement regime is no longer valid, because $\sigma\leq0$. For $R\lesssim30$\AA, figure \ref{figure_2} shows that the Stark shift, computed up to the zeroth order, is underestimated. The results become much more accurate, when first order terms are included, and seem efficient enough for describing QCSE in spherical semiconductor QDs. As soon as $R\gtrsim30$\AA, our results diverge from experimental data, as expected. When Coulomb interaction and polarization energy are included in the strong confinement regime, the weak field limit is no longer valid for QD radii 30\AA~$\lesssim R\lesssim50$\AA. Thus, a future work may focus on conciliating strong confinement regime and strong field limit. The case of the weak confinement regime is much more difficult to study, even in the weak field limit.
\section{Lamb shift} \label{sec_4}
The Lamb effect comes from the effect of a quantized electromagnetic field on the motion of a quantum particle of mass $m^*$ and charge $qe$ in a potential $U(\rr)$. Two popular methods to compute the resulting energy level shift are: the Bethe approach, which is a stationary second-order perturbation approach applied to a Pauli-Fierz Hamiltonian in the Coulomb gauge \cite{Bethe_1947}, and the Welton approach, which attributes the Lamb shift to particle position fluctuations \cite{Welton_1948}. They both lead to the same expression for the Lamb shift of a particle state $|\mathbf{n}\rangle$, $\Delta
E_\nn=\frac\alpha{3\pi}\frac{q^2}{m^{*2}}\log\!\!\left(\frac{m^*}\kappa\right)$ $\langle\nn|\nabla^2U(\rr)|\nn\rangle$, where $\alpha$ is the fine-structure constant, and $\kappa$ a IR cut-off, identified with $\langle|E_\mm-E_\nn|\rangle$ \cite{Bethe_1947, Welton_1948}.

For a particle of mass $m^*$ and charge $\pm e$, confined by the infinite spherical potential well $V(\rr)$ of a spherical QD, the Lamb shift of a level $E_{ln}$ turns out to be negative
  \begin{equation}
\Delta E_{ln}=-\frac{16\alpha}{3\pi}\frac{\lambdabarre^{*2}}{R^2}E_{ln}\log\!\!\left(\frac R{R^*_\mathrm{min}}\right)\!. \label{DE_lamb_general}
  \end{equation}
Here $\lambdabarre^*$ is the particle reduced Compton wavelength and $R^*_\textrm{\scriptsize min}=\frac\pi2\sqrt{\frac73}\lambdabarre^*$ is a QD radius lower bound, dictated by our non-relativistic assumption: for $R\leq R^*_\textrm{\scriptsize min}$, the particle confinement energy is of the order of its rest energy.

For a confined electron, its ground state undergoes a Lamb shift $\Delta E^\mathrm{Lamb}_{01}\approx-4.85~10^{-3}\mu$eV, for $R=1$nm. As $|\Delta E^\mathrm{Lamb}_{01}|$ decreases with $R$, it is unfortunately not observable in QDs of realistic radii. This is not so for an interactive electron-hole pair in a spherical semiconducting QDs, at least in the strong confinement regime.

The Lamb shift of the exciton ground state is made of four contributions. The first two contributions are due to the confinement infinite potential well $V(\rr_{\e,\h})$, given by equations (\ref{DE_lamb_general}), with $m^*_{\e,\h}$. The second is due to the Coulomb potential, and has the form of the Lamb shift observed in real atoms. Finally, the third comes from the pseudo-potential $W(\rr_{\e\h})$.

\begin{figure}
\caption{Lamb shift undergone by the electron (--$\!~$--), the hole (--$\!~\cdot~\!$--) and the exciton (---), when the exciton occupies its ground state in the strong confinement regime for $CdS_{0.12}Se_{0.88}$ microcrystallites.} \label{figure_3}
\vs{-.2cm}
    \begin{center}
\begin{picture}(0,0)%
\includegraphics{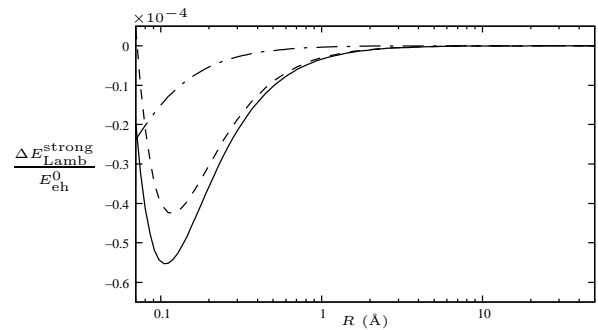}%
\end{picture}%
\setlength{\unitlength}{4144sp}%
\begingroup\makeatletter\ifx\SetFigFontNFSS\undefined%
\gdef\SetFigFontNFSS#1#2#3#4#5{%
  \reset@font\fontsize{#1}{#2pt}%
  \fontfamily{#3}\fontseries{#4}\fontshape{#5}%
  \selectfont}%
\fi\endgroup%
\begin{picture}(2967,1955)(-213,-990)
\put(1383,-944){\makebox(0,0)[b]{\smash{{\SetFigFontNFSS{5}{6.0}{\rmdefault}{\mddefault}{\updefault}{\color[rgb]{0,0,0}$R$ (\AA)}%
}}}}
\put(-198,-21){\makebox(0,0)[rb]{\smash{{\SetFigFontNFSS{5}{6.0}{\rmdefault}{\mddefault}{\updefault}{\color[rgb]{0,0,0}$\ds\frac{\Delta E^{\mathrm{strong}}_{\mathrm{Lamb}}}{E^0_{\e\h}}$}%
}}}}
\put(  1,866){\makebox(0,0)[lb]{\smash{{\SetFigFontNFSS{5}{6.0}{\rmdefault}{\mddefault}{\updefault}{\color[rgb]{0,0,0}$\times10^{-4}$}%
}}}}
\end{picture}%
  \end{center}
\vs{-.8cm}
\end{figure}

In the strong confinement regime, the Lamb shift undergone by the electron-hole pair ground state is evaluated, up to  $(\frac R{a^*})^2$ order, as $\Delta E^{\mathrm{strong}}_\mathrm{Lamb}=\Delta E^{\mathrm{strong}}_\e+\Delta E^{\mathrm{strong}}_\h$, where $\frac{\Delta E^{\mathrm{strong}}_{\e,\h}}{E_{\e\h}}=\frac{-16\alpha}{3\pi\varepsilon}\frac{\lambdabarre^{*2}_{\e,\h}}{R^2}\log\!\!\left(\!\frac R{R^{\e,\h}_\mathrm{min}}\!\right)\!\!\left[1-\!\left\{\frac{0.619\mu}{m^*_{\e,\h}}+\frac{2.111}{\pi^2}\right\}\!\frac R{a^*}\right.$
$\left.+\!\left\{\frac{0.699\mu}{m^*_{\e,\h}}-\frac{1.470}{\pi^2}+\frac83\right\}\!(\frac R{a^*})^2\!\right]\!$, $\lambdabarre^*_{\e,\h}$ being the electron and hole reduced Compton wavelengths and $R^{\e,\h}_{\textrm{\scriptsize min}}$ being minimal radii in the semiconductor QD. As $R\geq R^{\e,\h}_\textrm{\scriptsize min}$, this energy shift is negative. To the best of our knowledge, this is an outstanding new property, since in real atoms, the Lamb shift is always positive. Figure \ref{figure_3} suggests the possibility of measuring this Lamb shift in semiconducting QDs of realistic radii. Table \ref{table_2}.a and b confirms this, particularly in $InAs$ microcrystallites. Orders of magnitude involved are equivalent to those observed in hydrogen atom, and agree with experimental results, for which relative energy shifts, identified as Lamb shifts, smaller than 100$\mu$eV are measured \cite{Toda_1998, Demidenko_2004}.

\begin{table}
\caption{\label{table_2}Lamb shift undergone by the exciton ground state in the strong confinement regime in $CdS_{0.12}Se_{0.88}$ or $InAs$ microcrystals {\bf a.} for $R=10$\AA~and {\bf b.} for $R=30$\AA, and {\bf c.} in the weak confinement regime.} \vs{-.5cm}
  \begin{center}
{\footnotesize
    \begin{tabular}{ccccc}
\hline
 & \multirow{2}{1.9cm}{Semiconductor} & \multirow{2}{1.9cm}{$CdS_{0.12}Se_{0.88}$} & \multicolumn{2}{c}{$InAs$}
\\
 & &  & heavy hole & light hole
\\
\hline
{\bf a.} & $\Delta E^{\textrm{\tiny strong}}_\mathrm{Lamb}$ (meV) & -2,05.10$^{-5}$ & -9,49.10$^{-3}$ & -3,69.10$^{-6}$
\vs{.05cm}
\\
{\bf b.} & $\Delta E^{\textrm{\tiny strong}}_\mathrm{Lamb}$ (meV) & -2,11.10$^{-6}$ & -1,48.10$^{-4}$ & -5,94.10$^{-4}$
\vs{.05cm}
\\
{\bf c.} & $\Delta E^{\textrm{\tiny weak}}_\mathrm{Lamb}$ (meV) & 7,25.10$^{-6}$ & 2,69.10$^{-10}$ & 9,07.10$^{-8}$
\vs{.05cm}
\\
\hline
    \end{tabular}}
  \end{center}
\vs{-.7cm}
\end{table}

In the weak confinement regime, the Lamb shift of the exciton ground state, up to $(\frac{a^*}R)^{2}$ order, is $\Delta E^{\mathrm{weak}}_\mathrm{Lamb}=\Delta E^{\mathrm{weak}}_\e+\Delta E^{\mathrm{weak}}_\h$, where $\frac{\Delta E^{\mathrm{weak}}_{\e,\h}}{E^*}=\frac{8\alpha}{3\pi\varepsilon}\frac{\lambdabarre^{*2}_{\e,\h}}{a^{*2}}\log\!\!\left(\frac{m_{\e,\h}^*}{\kappa^*_{\e,\h}}\right)\!$. It is independent of the QD radius, and reveals the excitonic quasi-particle properties of the electron-hole pair in this regime. In the limit of infinite hole mass $\lambda\gg1$, only the electronic term $\Delta E^{\mathrm{weak}}_\e$ contributes, so that the Lamb shift undergone by the ground state of an hydrogen-like atom is retrieved. Following this analogy, the IR cut-offs should be both taken as $\kappa^*_{\e,\h}\approx19.8E^*$ \cite{Bethe}. Table \ref{table_2}.c suggests that the Lamb shift in the weak confinement regime is not observable at the moment.
\section{Purcell effect} \label{sec_5}
A two-level quantum atom, built from two states $|\nn\rangle$ and $|\mm\rangle$ of a quantum charged particle, of respective energies $E_\nn<E_\mm$, is fit into an electromagnetic cavity of quality factor $Q$, resonant at a frequency $\omega$ close to the Bohr frequency $\omega_\nn^\mm=E_\mm-E_\nn$. It interacts with a single dynamical confined electromagnetic cavity mode, also named quasi-mode, characterized by its effective volume $V_\mathrm{mode}$, and with a continuum of other external electromagnetic modes. In the electric dipole approximation and in a weak coupling regime, assuming that the two-level quantum system-confined mode coupling is treated as a perturbation to the confined mode-continuum coupling, the spontaneous emission rate of the radiative transition $|\mm\rangle\rightarrow|\nn\rangle$ is $A_\nn^\mm=2|\langle\mm|\ddd|\nn\rangle|^2\frac Q{V_\mathrm{mode}}$, $\ddd$ being the particle electric dipole moment. It is given by the Fermi golden rule applied to the Jaynes-Cummings Hamiltonian \cite{Jaynes_1963}. From  the definition $A_\nn^\mm=F_\nn^\mm\!~^0\!\!A_\nn^\mm$ and from the spontaneous emission rate $^0\!\!A_\nn^\mm=\frac{(\omega_\nn^\mm)^3}{3\pi}|\langle\mm|\ddd|\nn\rangle|^2$ in absence of electromagnetic cavity  \cite{Ballentine}, the spontaneous emission rate is enhanced by the Purcell factor $F_\nn^\mm=\frac{3Q}{4\pi^2}\frac{(\lambda_\nn^\mm)^3}{V_\mathrm{mode}}$, where $\lambda_\nn^\mm$ is the wavelength associated to $\omega_\nn^\mm$.

If the emitted photon escapes the two-level quantum system and the electromagnetic cavity, without being re-absorbed, the confined mode-continuum coupling dominates. As the Purcell effect competes with the adverse working of Rabi oscillations \cite{Tannoudji}, a sufficient condition for the validity of the weak coupling is $\tau_\nn^\mm\Omega_\nn^\mm\ll1$. Here, $\tau_\nn^\mm=\frac Q{\omega_\nn^\mm}$ is defined as the photon relaxation time and $\Omega_\nn^\mm=\sqrt{\frac\omega{2V_{\textrm{\tiny mode}}}}|\langle\mm|\ddd|\nn\rangle|$ the Rabi frequency. As $\tau_\nn^\mm\Omega_\nn^\mm$ scales as $\propto Q$, this imposes an upper bound on $Q$. The higher $Q$ is, the smaller is the resonance disagreement $|\omega-\omega_\nn^\mm|$, which is responsible for the Rabi oscillations evanescence. Therefore, Rabi oscillations can be sustained in the cavity, inhibiting the Purcell effect \cite{Brune_1996}.

In practice, the effective cavity mode volume $V_{\textrm{\scriptsize mode}}$  is experimentally measured. A review of electromagnetic cavities of quality factor $Q\geq2000$, characterized by an effective volume $V_\mathrm{mode}\approx\beta(\lambda_\nn^\mm)^3$, $\beta$ being of order unity, is given in \cite{Vahala_2003}. For such cavities, the Purcell factor becomes independent from the radiative transition, and $F=\frac{3Q}{20\pi^2}$ if $\beta=5$. Then, to observe the Purcell effect, the quality factor should be larger than the lower bound $Q_{\textrm{\scriptsize min}}\approx66$, which is the case in common electromagnetic cavities.

In a semiconductor QD, as the Purcell effect concerns radiative transitions  between any two QD eigenstates, the confined hole is assumed to be in its ground state. According to section {\bf \ref{sec_2}}, electron tunneling should be discarded, and only electronic radiative transitions involving energy levels lower than the maximum amplitude of the real finite potential step should be considered. These considerations should be valid in the strong confinement regime. Spontaneous emission rates with or without electromagnetic cavity are analytically computed between two confined electron states of energies $E_{ln}<E_{l'n'}$ as $^0\!\!A_{lnm}^{l'n'm'}=\frac{64\sqrt\varepsilon\alpha}3\! \left[\frac{2\pi}{\lambda_{ln}^{l'n'}}\right]^{\!3}\!\!(I^{ll'}_{nn'})^2J_{ll'}^{mm'}R^2$ and $A_{lnm}^{l'n'm'}=F~^0\!\!A_{lnm}^{l'n'm'}$. Here, $I^{ll'}_{nn'}$ and $J_{ll'}^{mm'}$ are exactly computable integrals. Since $^0\!\!A^{l'n'm'}_{lnm}\propto R^{-4}$, spontaneous emission rates are large for small QDs, illustrating a typical quantum behavior.

\begin{figure}
\caption{Diagram of states and transitions in the three-level {\it red} QD LASER.  The pumping is realized between a lowest level $|010\rangle$ of the spherical QD spectrum (the {\it ground state level}) and a level $|090\rangle$ (the {\it excited state}) higher than the highest level $|180\rangle$ of the LASER transition (the {\it intermediate state}).} \label{trois_niveaux} \vs{-.2cm}
    \begin{center}
\begin{picture}(0,0)%
\includegraphics{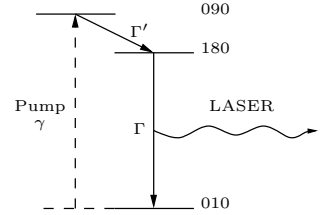}%
\end{picture}%
\setlength{\unitlength}{4144sp}%
\begingroup\makeatletter\ifx\SetFigFontNFSS\undefined%
\gdef\SetFigFontNFSS#1#2#3#4#5{%
  \reset@font\fontsize{#1}{#2pt}%
  \fontfamily{#3}\fontseries{#4}\fontshape{#5}%
  \selectfont}%
\fi\endgroup%
\begin{picture}(1813,1250)(4226,-3871)
\put(5332,-2927){\makebox(0,0)[lb]{\smash{{\SetFigFontNFSS{6}{7.2}{\rmdefault}{\mddefault}{\updefault}{\color[rgb]{0,0,0}180}%
}}}}
\put(5332,-2696){\makebox(0,0)[lb]{\smash{{\SetFigFontNFSS{6}{7.2}{\rmdefault}{\mddefault}{\updefault}{\color[rgb]{0,0,0}090}%
}}}}
\put(5006,-3414){\makebox(0,0)[rb]{\smash{{\SetFigFontNFSS{6}{7.2}{\rmdefault}{\mddefault}{\updefault}{\color[rgb]{0,0,0}$\Gamma$}%
}}}}
\put(4914,-2835){\makebox(0,0)[lb]{\smash{{\SetFigFontNFSS{6}{7.2}{\rmdefault}{\mddefault}{\updefault}{\color[rgb]{0,0,0}$\Gamma'$}%
}}}}
\put(4380,-3275){\makebox(0,0)[b]{\smash{{\SetFigFontNFSS{6}{7.2}{\rmdefault}{\mddefault}{\updefault}{\color[rgb]{0,0,0}Pump}%
}}}}
\put(5332,-3856){\makebox(0,0)[lb]{\smash{{\SetFigFontNFSS{6}{7.2}{\rmdefault}{\mddefault}{\updefault}{\color[rgb]{0,0,0}010}%
}}}}
\put(5381,-3272){\makebox(0,0)[lb]{\smash{{\SetFigFontNFSS{6}{7.2}{\rmdefault}{\mddefault}{\updefault}{\color[rgb]{0,0,0}LASER}%
}}}}
\put(4380,-3369){\makebox(0,0)[b]{\smash{{\SetFigFontNFSS{6}{7.2}{\rmdefault}{\mddefault}{\updefault}{\color[rgb]{0,0,0}$\gamma$}%
}}}}
\end{picture}%
    \end{center}
\vs{-.8cm}
  \end{figure}

In the strong confinement regime, we discuss now the possibility of exploiting Purcell effect to produce red-light LASER emission at $755$nm from $InAs$-spherical QDs of radius $R=25$nm \cite{Lee_2002}. Figure \ref{trois_niveaux} describes the lasing mechanism, and table \ref{table} collects required numerical data. Spontaneous emission is the only relevant phenomena: stimulated emission and absorption should be discarded, and non-radiative effects should be omitted.

  \begin{table}
\caption{Table of wavelengths and spontaneous emission rates of the three-level QD {\it red} LASER presented at \ref{trois_niveaux} in comparison with the LASER transition of $He$-$Ne$ LASER.}
\label{table} \vs{-.2cm}
    \begin{center}
{\footnotesize
      \begin{tabular}{ccccc}
\hline
\multirow{2}{1cm}{LASER} & \multirow{2}{1cm}{Transition} & wavelength & $^0\!\!A_{lnm}^{l'n'm'}$
\\
 & & (nm) & (MHz)
\\
\hline
\multirow{2}{1.7cm}{QD LASER} & $|090\rangle\!\rightarrow|180\rangle$ & 6.02~10$^3$ & 401
\\
 & $|180\rangle\!\rightarrow|010\rangle$ & 755 & 0.617
\\
\hline
 $He$-$Ne$ & LASER & 632 & $\approx50$
\\
\hline
      \end{tabular}}
    \end{center}
\vs{-.7cm}
  \end{table}

Assuming that the decay $|090\rangle\rightarrow|180\rangle$, of relaxation rate $\Gamma'$, is faster than the decay $|180\rangle\rightarrow|010\rangle$ of relaxation rate $\Gamma$: the intermediate state is metastable and $\Gamma'\gg\Gamma$. In a stationary regime with weak pumping $\omega\ll\Gamma'$, the population inversion holds only for $\omega\geq\Gamma$. This implies that $\Gamma'\gg\Gamma$, which means that the excited state is almost empty, and the intermediate state is the most populated. As $\Gamma'\geq\!\!~^0\!\!A_{180}^{090}\approx401$MHz and $\Gamma=A^{180}_{010}\approx9.38Q$kHz, the  assumptions for having a red-light three-level LASER are met, if $F\approx100$, {\it i.e.} if $Q\approx6500$, and if $\omega$ is of the order of $80$MHz. In particular, the condition for obtaining the Purcell effect is met $\tau_{010}^{180}\Omega_{010}^{180}\approx6,3.10^{-3}$. Table \ref{table} suggests that spontaneous emission rates are of the same order of magnitude than those of the $He$-$Ne$ LASER. So, the Purcell effect coupled to the QDs artificially tailored spectrum leads to the possibility of producing visible light LASER emission. However, even in the strong confinement regime, this result is to be considered with care since electron-hole Coulomb interaction is not yet included.
\section{Conclusion}
In this paper, a novel approach to some interesting properties of spherical semiconducting QDs is presented. It is based on an improved EMA classical model with an added effective pseudo-potential. This allows extensive analytic calculations of physical quantities yielding considerably better agreement with numerical or/and measured data for Quantum Size Effects and Quantum Confinement Stark Effects. Using this model, the Lamb shift in spherical semiconductor QDs is computed. It turns out to be negative and in principle observable, at least in the strong confinement regime. Finally our study also demonstrates the utility of the Purcell effect, predicted for atoms, for QD-LASER emission in the visible. These wide ranging theoretical results are encouraging for further investigation of QDs structure, based on this improved EMA model.

%
%
%

\end{document}